# A Memory Bandwidth-Efficient Hybrid Radix Sort on GPUs

Elias Stehle
Technical University of Munich (TUM)
Boltzmannstr. 3
85748 Garching, Germany
stehle@in.tum.de

Hans-Arno Jacobsen
Technical University of Munich (TUM)
Boltzmannstr. 3
85748 Garching, Germany
jacobsen@in.tum.de

## ABSTRACT

Sorting is at the core of many database operations, such as index creation, sort-merge joins, and user-requested output sorting. As GPUs are emerging as a promising platform to accelerate various operations, sorting on GPUs becomes a viable endeavour. Over the past few years, several improvements have been proposed for sorting on GPUs, leading to the first radix sort implementations that achieve a sorting rate of over one billion 32-bit keys per second. Yet, state-of-the-art approaches are heavily memory bandwidth-bound, as they require substantially more memory transfers than their CPU-based counterparts. Our work proposes a novel approach that almost halves the amount of memory transfers and, therefore, considerably lifts the memory bandwidth limitation. Being able to sort two gigabytes of eight-byte records in as little as 50 milliseconds, our approach achieves a 2.32-fold improvement over the state-of-the-art GPU-based radix sort for uniform distributions, sustaining a minimum speed-up of no less than a factor of 1.66 for skewed distributions. To address inputs that either do not reside on the GPU or exceed the available device memory, we build on our efficient GPU sorting approach with a pipelined heterogeneous sorting algorithm that mitigates the overhead associated with PCIe data transfers. Comparing the end-to-end sorting performance to the state-of-the-art CPU-based radix sort running 16 threads, our heterogeneous approach achieves a 2.06-fold and a 1.53-fold improvement for sorting 64 GB key-value pairs with a skewed and a uniform distribution, respectively.

## 1. INTRODUCTION

Many of today's database systems are facing unprecedented loads as they must cope with data that is generated by hundreds of millions of people, devices, and sensors [7, 9]. Analysing, filtering, and querying the enormous amount of data in a timely manner becomes increasingly difficult. In an endeavour to keep systems responsive, a lot of effort is put into adapting database systems to modern hardware trends [21, 6, 23, 1, 3, 33, 5, 22, 30, 36]. The availability of low-cost memory, for instance, has given rise to the wide adoption of in-memory databases [35, 26, 24, 8]. In many cases, this has shifted the bottleneck from I/O to memory bandwidth and compute performance. Moreover, the rise of multi-core architectures, vector processing capabilities, and growing cache sizes requires to rethink many parts of database systems.

Sorting is no exception to this effort. As a fundamental operation in database systems, sorting finds its application in index creation, user-requested output sorting, and sort-merge joins [13]. Moreover, sorting can speed up duplicate removal, ranking, and grouping operations [13]. Therefore, a lot of research has been devoted to identifying efficient sorting algorithms that utilise modern hardware features and scale well across multiple cores, processors, and even nodes [21, 6, 35, 40, 24, 33, 22, 8]. After having recently achieved sorting rates of over one billion keys per second [28], Graphics Processing Units (GPUs), featuring thousands of cores and a memory bandwidth of several hundred gigabytes per second, emerged as a promising platform to accelerate sorting. Besides approaches that are based on sorting networks [25, 12], merge sort [37, 34, 35], and sample sort [27, 11], the most promising results for larger problem sizes have been shown for implementations using a radix sort [18, 16, 34, 35, 28].

A major challenge arising when trying to make use of the massive parallelism of GPUs for sorting is the fact that a key's position within the output sequence depends on all other keys. Previous work has addressed this issue by using a least-significant-digit-first radix sort (LSD radix sort) that iterates over the keys' bits from the least-significant to the most-significant digit, considering an implementation specific number of consecutive bits at a time. With each sorting pass, a stable counting sort is used to partition the keys into buckets according to the bits being considered with the current pass [16, 34, 35, 28]. The stable counting sort computes each key's offset by counting the number of keys with a smaller digit value and, as it needs to be stable, the keys with the same digit value preceding the key in the input sequence. To achieve concurrency, GPU-based implementations split the input into a sequence of small blocks (a few thousand keys) that are processed in parallel. For each block, a local histogram over the keys' digit values is computed, and the prefix-sum over these histograms is used to determine a key's position within the output sequence. Since the whole input has to be read twice and written once with each sorting pass, radix sort implementations aim to increase the number of bits being considered with each sorting pass, in order to lower the number of passes and the amount of memory transfers. However, as the size of the histogram grows exponentially with the number of bits being considered with each sorting pass, the growing complexity of the prefix-sum computation and the small on-chip memory impose a limit on the number of bits per digit. Due to these limitations, state-of-the-art approaches are restricted to consider only five bits at a time. Incurring a considerable amount of memory transfers, such as reading or writing the input 39 times in the case of 64-bit keys, the sorting rate is ultimately bound by the available memory bandwidth.

In order to lift the memory bandwidth limitation, this work presents a novel, hybrid radix sort that is able to efficiently sort on eight bits with each pass. This reduces the number of sorting passes and therefore the total amount of memory transfers by a factor of at least 1.6. In contrast to an LSD radix sort that is used by state-of-the-art implementations (e.g., CUB), the presented approach does not rely on stable sorting passes [29]. Therefore, it is not restricted to respecting the order of preceding sorting passes for keys falling into the same bucket. Lifting this constraint enables our approach to use native shared memory atomic operations that became available with recent GPU microarchitectures to mitigate the downside of considering more bits with each sorting pass [31, 32]. Our hybrid approach starts from the most-significant bit and proceeds towards the least-significant bit, partitioning the keys into smaller and smaller buckets. It avoids running into situations where the partitioning of the input into too many buckets would negatively impact performance, by finishing with a local sort as soon as a bucket is small enough to fit into on-chip memory. As the local sort performs the sorting in on-chip memory, it needs to access the device memory only twice, once for reading and once for writing the keys, no matter how many sorting passes it requires. This further saves essential memory bandwidth and boosts performance for favourable distributions. While a typical parallel most-significant-digit-first radix sort (MSD radix sort) may incur load balancing issues for skewed distributions that result in buckets of greatly varying size, we efficiently utilise the low-overhead scheduling mechanisms of the GPU to avoid any load imbalance, by subdividing every bucket into tiny, fixed-size blocks that can be evenly distributed amongst the GPU's Streaming Multiprocessors (SMs).

To circumvent the overhead associated with a large number of kernel invocations, we use only a constant number of invocations during each sorting pass. Rather than using at least one invocation for each bucket, passing the memory offset of the bucket's keys and its size as arguments, we generate that information as a byproduct of a sorting pass and place it in device memory, from where it can be read in the subsequent pass to determine the work assignments. Moreover, we show that the use of shared memory atomic operations is highly efficient for almost any key distribution and introduce measures to mitigate performance degradation for highly skewed distributions.

In order to address inputs that either do not reside on the GPU or exceed the available device memory, we present a pipelined heterogeneous sorting algorithm that mitigates the overhead associated with PCIe data transfers. By splitting the input into multiple sub-problems, we are able to interleave several processing stages, allowing us to exploit the full-duplex capability of the PCIe bus while simultaneously sorting on the GPU. In order to max out the limited device memory, we propose an in-place replacement strategy that immediately reuses memory by returning a sorted run while concurrently replacing the contents with the next sub-problem. This allows us to support larger sub-problems, which improves the overall performance for sorting large inputs of tens of gigabytes.

We evaluate the hybrid radix sort for various key and value sizes over twelve different, increasingly skewed distributions and compare it to the state-of-the-art GPU-based radix sort (CUB)[29]. Our experimental results demonstrate that the hybrid radix sort efficiently capitalises on the 1.6-fold reduction in the amount of memory transfers, seeing no less than a 1.58-fold improvement over CUB. Being able to sort two gigabytes of 64-bit keys with an associated 64-bit value in as little as 56 milliseconds, our approach peaks out at a four-fold speed-up. Building on the results of our hybrid radix sort, we evaluate the end-to-end performance for our heterogeneous sort and compare it to the state-of-the-art CPU-based radix sort running 16 threads [8]. Being able to sort 16 GB comprised of key-value pairs with a skewed distribution in as little as 3.37 seconds, the heterogeneous sort outperforms PARADIS by a factor of 2.64 [8]. Sorting an input of 64 GB with a skewed distribution, we still see a 2.06-fold improvement over PARADIS, despite the fact that our CPU-side processing on a weaker processor with only six-cores contributes more than 9.3 seconds to the 16 second total.

Overall, the contributions of this work are five-fold:

1. We present a novel, hybrid radix sort for GPUs that proceeds from the most-significant to the least-significant bit to circumvent the downside of considering more bits with each sorting pass. Not relying on stable sorting passes allows our approach to efficiently sort on eight bits at a time, and therefore reduce the number of passes and the amount of memory transfers by no less than a factor of 1.6.
2. We successfully address the challenges arising from implementing an MSD-based radix sort on GPUs, such as load balancing and congestion issues for skewed distributions and performance degradation due to bucket handling.
3. Using a local sort for sorting small buckets, we are not only able to avoid running into situations with an overwhelmingly large number of buckets, but also to considerably boost the performance for favourable key distributions, culminating in a four-fold speed-up.
4. As an MSD-based radix sort may result in millions of buckets that need to be kept track of, we establish an analytical model that is used to calculate the upper bounds on the number of buckets and analyse the memory requirements. The model shows the feasibility of our hybrid radix sort, indicating that the additional memory overhead, such as for keeping track of buckets, does not exceed a mere 5% of an LSD radix sort.
5. We address inputs that either do not reside on the GPU or exceed the available device memory using a pipelined heterogeneous sorting algorithm that mitigates the overhead associated with PCIe data transfers. In order to efficiently exploit the limited device memory, we propose an in-place replacement strategy that improves the overall performance for large inputs.

Table 1: Notation

| symbol | description |
|---|---|
| $k$ | number of bits per key |
| $d$ | number of bits per digit |
| $KPT$ | number of keys per thread |
| $KPB$ | number of keys per block |
| $\hat{\partial}$ | threshold for local sorting |
| $\underline{\partial}$ | threshold for merging buckets |

This paper is organised as follows. In Section 2, we introduce the basics of radix sorting and present the fundamental concepts of general-purpose computing on GPUs. Section 3 analyses the state-of-the-art approaches for sorting on GPUs. Section 4 presents the hybrid radix sort, how it is realised and how performance drops are mitigated. Section 5 addresses our heterogeneous sorting algorithm that aims to mitigate the overhead introduced with PCIe data transfers. Section 6 evaluates the performance of the presented approach and compares it to the state-of-the-art.

## 2. BACKGROUND

This section gives a quick introduction to radix sorting followed by an overview of recent GPU microarchitectures. This work focuses on NVIDIA GPUs and the *CUDA* computing platform. CUDA has been widely adopted for general purpose computing on GPUs and allows to tailor implementations to specific hardware characteristics. The notation used throughout this work is presented in Table 1.

### 2.1 Radix Sorting

Radix sorting relies on the reinterpretation of a $k$-bit key as a sequence of $d$-bit digits, which are considered one at a time. The basic idea is, that splitting the $k$ bits of the keys into smaller $d$-bit digits results in a small enough radix $r = 2^d$, such that the keys can efficiently be partitioned into $r$ distinct buckets. As sorting on each digit can be done with an effort that is linear in the number of keys $n$, the whole sorting can be achieved with a total complexity of $\mathcal{O}(\lceil k/d \rceil \times n)$. Iterating over the keys' digits can be performed in two fundamentally different ways. Either by proceeding from the most-significant to the least-significant digit (MSD radix sort), or vice versa (LSD radix sort).

The MSD radix sort starts with the most-significant digit and partitions the keys into a sequence of $r$ distinct buckets, according to their digit value. This can be done using a counting sort, which starts computing the histogram over the keys' most-significant digit. As the histogram reflects the number of keys that shall be put into each of the $r$ buckets, computing the exclusive prefix-sum over these counts yields the memory offsets for each of the buckets. Finally, the keys are scattered into the buckets according to their digit value. Recursively repeating these steps on subsequent digits for the resulting buckets ultimately yields the sorted sequence.

In contrast, the LSD radix sort starts with the least-significant digit and performs a stable sort in subsequent passes. That is, if there is a tie on the digit's value of any two keys, the original order of the preceding pass is preserved. Hence, during a sorting pass, a key's position is given by the number of keys with a lower digit value plus the number of keys that have the same digit value and precede the key in the input sequence.

### 2.2 GPU Architecture

GPU architectures have been steadily scaling up their core counts over time, proliferating in thousands of simple cores today. Moreover, discrete GPUs feature their own device memory that provides transfer rates of up to 750 GB/s [32]. The basic building block of a GPU is a SM. Each SM consists of a set of cores (e.g., 64, 128, or 192), a *register file*, *shared memory*, and an L1 cache. The register file is used to hold the registers of all threads that reside on an SM. An important limitation of registers is that they cannot be addressed dynamically. Hence, declaring an array and accessing it based on an index that cannot be resolved at compile time, would render the use of registers impossible. In contrast, shared memory is dynamically addressable and shared by a whole group of threads, referred to as *thread block*. A thread block is the atomic unit that is scheduled on an SM. It is defined by the amount of shared memory that is required, a function (the kernel), and the number of threads that execute the given function. It is possible, and even desired, that several thread blocks reside on an SM at any given time, increasing the *occupancy*. For every thread block that resides on an SM, the required number of registers and the amount of shared memory is allocated to the thread block. Thus, the maximum number of blocks that can possibly reside on a single SM is implied by the resources a thread block requires and the resources that are available on an SM. For example, an SM with 96 KB of shared memory and 65 536 registers, could accommodate up to eight thread blocks of 256 threads, if each block requires eight KB of shared memory and 16 registers per thread (a total of 4 096 registers per block). Each thread block is subdivided into a set of warps, currently comprising 32 threads. All threads of a warp are executed in a lockstep manner. With several thread blocks and therefore several warps residing on an SM, the scheduler can choose from the set of resident warps that are ready for being executed rather than waiting for a single warp to get ready (e.g., for hiding latency from memory accesses).

## 3. RELATED WORK

Over the years many different approaches have been pursued for sorting on GPUs. Kipfer et al. have proposed a solution that is based on the odd-even sorting network and an approach using a bitonic merge sort algorithm [25]. *GPUTeraSort*, introduced by Govidaraju et al., aims to address larger keys as well as larger problem sizes that previously have been limited to the GPU's device memory [12]. Moreover, they used an index sort that uses the CPU to rearrange the key-value pairs based on the key-index pairs that are sorted and returned by the GPU. To reduce the overall complexity of a sorting network-based approach, which exhibits a complexity of $\mathcal{O}\left(n \log^2 n\right)$, Harris et al. propose a solution that divides the input sequence into smaller subsequences, sorts them locally using a binary radix sort, i.e., a radix of two, and merges the chunks using a parallel bitonic sort [17]. Similarly, Ye et al. proposed *Warpsort*, which sorts the chunks using a bitonic sorting network [41]. In addition, their approach avoids costly synchronisation by exploiting the synchronous execution of a warp's threads. Other merge-based approaches have been presented by Satish et al. [34, 35], Davidson et al. [10], Green et al. [15], and Tansic et al. [38].

Apart from merge-based approaches, promising results were shown for implementations building on a distribution-based sort, such as a radix sort. As part of their introduction of a multi-pass scatter operation that aims to coalesce memory writes, He et al. present an MSD radix sort that uses a fixed number of partitioning passes [18]. The MSD radix sort partitions the input, considering five bits at a time. After performing a fixed number of partitioning passes, a bitonic sort is used to sort each of the partitions. The approach works for a uniform distribution, which is assumed when the fixed number of required partitioning passes is calculated. For skewed distributions, however, their sort would not gain a big advantage from the partitioning passes. For instance, assuming an input that, according to the algorithm's logic, would be considered for two partitioning passes. If the keys' bits are all zero on their most-significant ten bits, the algorithm would spend time on the two partitioning passes, while it still ends up with one single partition. Sintorn et al. present a hybrid approach that starts with a partitioning pass, using either a quicksort or a bucket sort, before sorting each of the resulting partitions with a merge sort [37]. The bucket sort uses an initial set of heuristic splitters, counts the keys belonging to each of the partitions defined by the splitters, and, if required, refines the splitters. Once the splitters have been examined, the keys are scattered into 1 024 partitions, which, in turn, are sorted using the proposed merge sort.

Satish et al., as well as Ha et al., propose an LSD radix sort, which coalesces writes to device memory by performing the key scattering in the local shared memory, prior to writing the local partitions to device memory [34, 16]. While Ha et al. sort on only two bits at a time, Satish et al. manage to use digits of four bits by repeatedly using a binary split within shared memory on each single bit, before writing the partitions to device memory. Satish et al. provide a thorough evaluation of comparison and non-comparison sorts on different architectures [35]. They examined that their radix sort, which is based on the approach presented by Satish et al. [34], is compute-bound, and make a case for their merge sort. To avoid the computational effort associated with the binary split, and save the amount of data being transferred, Merrill et al. present a tuned radix sort that achieves a sorting rate of over one billion 32-bit keys per second, yet, reaches its optimum for sorting four-bit digits [28]. The approach of Merrill et al. has been integrated into the CUB header library, which is developed and maintained by NVIDIA Research [29]. As part of CUB, the radix sort is able to efficiently sort on five bits at a time.

## 4. ON-GPU HYBRID RADIX SORTING

This section describes our approach to radix sorting on GPUs. We give an overview of our sorting algorithm, introduce its two fundamental components, the *counting sort* and the *local sort*, and explain how we designed the hybrid radix sort for GPUs. We first limit the presentation of the approach to the sorting of unsigned integer keys before explaining how it can be extended to sort keys and key-value pairs of any primitive data type (e.g., *int*, *float*, *double*).

### 4.1 The Hybrid Radix Sort

The proposed algorithm is based on an MSD radix sort, which recursively partitions the keys into smaller and smaller buckets until the buckets are eventually small enough to be sorted in on-chip shared memory. We distinguish between a *counting sort*, which performs the aforementioned partitioning of a bucket into sub-buckets, and a *local sort*, that brings all keys of a small bucket into sorted order. The algorithm starts with a counting sort on the most-significant digit (the $d$ most-significant bits) and produces a sequence of $r = 2^d$ sub-buckets, each containing a partition of the keys that share the same value on their most-significant digit. With every subsequent sorting pass, each sub-bucket that resulted from the partitioning of the buckets in the preceding pass is either further partitioned using another counting sort, or sorted using a local sort. While proceeding to the next sorting pass, the digit according to which the counting sort partitions the buckets into sub-buckets is advanced by one towards the least-significant digit. The algorithm is finished once all keys are sorted up to and including the least-significant digit, or, if all buckets have been sorted with a local sort. The general workflow is illustrated in Figure 1. It depicts a local sort as a waved arrow pointing from a single bucket to a location in memory for the sorted output, and a counting sort as a set of arrows that point from a single bucket to a sequence of sub-buckets.

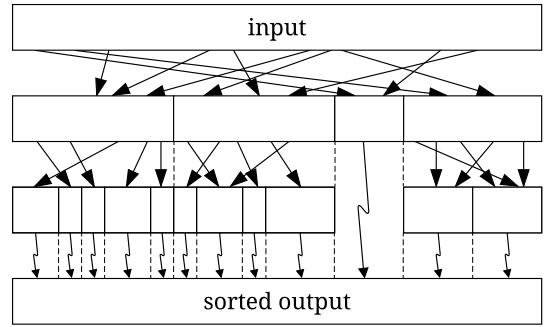

**Figure 1: The hybrid radix sort**

While the local sort works in-place, the counting sort requires auxiliary memory to which the partitioned keys are written. In order to reuse memory, we are using double-buffering for the whole sorting algorithm. With every sorting pass, memory for the input and the output is exchanged, such that the memory for the output of the preceding pass becomes the input of the current pass, and the previous pass's input memory is reused for the output. As the memory for the input and the output is alternating with each pass, we return the final sorted sequence within the memory of the original input if the number of digits, $\lceil k/d \rceil$, is even, and within the auxiliary memory otherwise. Since the algorithm might finish early, i.e., if all buckets have been sorted using a local sort prior to reaching the least-significant digit, we make sure that a local sort always places the sorted key sequence in the memory being used to return the final sorted output.

As the local sort is sorting a bucket's keys within on-chip shared memory, it is limited to sort a maximum of $\hat{\partial}$ keys, which is implied by the key size and the available hardware resources. To take advantage of the fact that preceding counting sort passes have already sorted the bucket's keys up to a certain digit, we can tune an LSD radix sort to only sort on the remaining digits.

Buckets that exceed the local sort threshold, $\hat{\partial}$, are partitioned into sub-buckets using a counting sort. The counting sort reads the keys starting at the bucket's offset from the input memory, partitions them into sub-buckets according to the specified digit and writes the sequence of sub-buckets cohesively into the output memory, such that the sub-bucket holding the keys with the smallest digit value starts at the same offset as the input bucket. An implementation of a counting sort for a single bucket follows these steps:

(1) *Compute the histogram over the digit values of all keys in the bucket to determine the size of each sub-bucket.*

(2) *Compute the exclusive prefix-sum over the histogram to get the offset for each of the r sub-buckets.*

(3) *Scatter the keys into the sub-buckets according to the keys' digit values.*

Table 2: Hybrid radix sorting example: sorting 16 keys of k=4 bits with d=2 bits and a radix of r=4

| | 0 | 1 | 2 | 3 | 4 | 5 | 6 | 7 | 8 | 9 | 10 | 11 | 12 | 13 | 14 | 15 |
|---|---|---|---|---|---|---|---|---|---|---|---|---|---|---|---|---|
| keys (radix 4) | 31 | 12 | 01 | 23 | 12 | 22 | 12 | 00 | 11 | 10 | 10 | 31 | 03 | 13 | 12 | 03 |
| histogram | 4 | 8 | 2 | 2 | | | | | | | | | | | | |
| prefix-sum | 0 | 4 | 12 | 14 | | | | | | | | | | | | |
| sort (radix 4) | bucket 0 | | | | bucket 1 | | | | | | | | bucket 2 | | bucket 3 | |
| | 01 | 00 | 03 | 03 | 12 | 12 | 12 | 11 | 10 | 10 | 13 | 12 | 23 | 22 | 31 | 31 |
| histogram | 1 | 1 | 0 | 2 | 2 | 1 | 4 | 1 | | | | | local | | local | |
| prefix-sum | 0 | 1 | 2 | 2 | 0 | 2 | 3 | 7 | | | | | local | | local | |
| sort (radix 4) | $b_0$ | $b_1$ | $b_3$ | | $b_0$ | $b_1$ | $b_2$ | | | | | $b_3$ | local | | local | |
| | 00 | 01 | 03 | 03 | 10 | 10 | 11 | 12 | 12 | 12 | 12 | 13 | 22 | 23 | 31 | 31 |

The presented approach is exemplified in Table 2, which shows the algorithm for 16 keys of a length of four bits. The radix sort is performed using two-bit digits with a radix of $r = 4$, requiring exactly two passes to fully sort the keys. The keys are represented in a base four notation. In the example, we set the threshold for local sorting to $\hat{\partial} = 3$, turning to a local sort for buckets of three keys or less.

## 4.2 Fine-Grained Parallelism on GPUs

While the presented algorithm allows to process individual buckets in parallel, the level of parallelism may not suffice to have enough threads in flight to hide the latency from memory accesses. Therefore, we introduce a higher degree of concurrency for the counting sort by splitting the $n$ keys of each bucket into a sequence of $\lceil n/KPB \rceil$ *key blocks*, each comprised of up to $KPB$ keys. Each key block is processed once during the computation of the histogram and once during the scattering step.

In order to decrease the overhead associated with kernel invocations, we use only a constant number of invocations per sorting pass, independent of the number of buckets being sorted. A kernel invocation instructs the GPU to execute a given kernel (function) by a specified number of thread blocks, each comprised of a given number of threads. Rather than adjusting the arguments (e.g., pointer to a bucket's keys, number of keys) for each bucket individually, using multiple invocations, we put that information into device memory as a byproduct of the prefix-sum computation and launch just enough thread blocks to have one for each key block of each bucket. During the computation of the histogram and the key scattering step, each thread block looks up the bucket and the block of keys it is assigned to by reading that information from device memory.

We proceed similarly for the local sort, where we assign exactly one thread block to each bucket. However, there is a downside to using only a single kernel invocation for all buckets that are sorted using a local sort. That is, there are just as many threads being assigned for processing a large bucket that has close to $\hat{\partial}$ keys, as there are for sorting a relatively small bucket of only a few keys. Thus, with many threads being over-provisioned for small buckets, this introduces additional overhead. We address this issue in two ways.

Firstly, we start merging tiny neighbouring sub-buckets whose total number of keys falls below a certain threshold $\underline{\partial}$. That is, after a counting sort has partitioned a bucket into $r$ sub-buckets, we merge any sequence of sub-buckets as long as their total number of keys is less than $\underline{\partial}$, with $\underline{\partial} \leq \hat{\partial}$. This further reduces the upper bound on the total number of buckets and avoids having too many tiny buckets, for which the scheduling of an own thread block would introduce considerable overhead, compared to the time that is spent on the sorting.

Secondly, instead of using a single kernel invocation that sorts all buckets whose size falls into the interval $[1, \hat{\partial}]$, we distinguish between different bucket sizes in that interval, e.g., bucket sizes of $[1, 128], (128, 256], (256, 512], ..., (..., \hat{\partial}]$ keys, respectively. For each of these subintervals, a kernel is invoked with each thread block provisioning just enough threads to process the respective number of keys. We refer to each of these as a *local sort configuration*, which represents the combination of a kernel, a number of threads per thread block, and the supported bucket size. In addition to adjusting the number of threads per thread block, this allows to specify a certain kernel that is optimised for sorting the given number of keys. Hence, for small buckets, a configuration with a sorting network or another comparison-based sorting algorithm could be devoted, turning to an LSD radix sort for configurations supporting buckets of a larger size.

## 4.3 Histogram

One of the key advantages of the proposed approach is, that, in contrast to an LSD radix sort, the hybrid radix sort does not rely on stable sorting passes. Therefore, it is not restricted to respecting the order of preceding sorting passes for keys falling into the same sub-bucket. Lifting this constraint enables our approach to use native shared memory atomic operations for the histogram computation and the key scattering step to mitigate the downside of considering more bits with each sorting pass.

Our histogram computation aggregates one histogram per block in shared memory. Every thread reads $KPT$ keys from device memory, iterates over them, and uses an *atomicAdd* operation to increment the counter in shared memory for the respective digit value. Once all threads of a block are done, the histogram that has been accumulated in shared memory is added to the global histogram by adding the respective counters in device memory.

Since all threads of a thread block share the same counters for the local histogram, highly skewed distributions with only few digit values potentially degrade the performance, as this causes all threads to simultaneously access the same counters in shared memory. In order to be able to max out the available memory bandwidth, each SM must achieve a processing rate of $\frac{8 \times BW}{k \times |SMs|}$ keys per second, where $BW$ denotes the peak memory bandwidth in bytes per second and $|SMs|$ the number of available SMs. Based on the number of SMs and the theoretical peak memory bandwidth of recent GPUs, this gives a required throughput of $3-4.5$ billion 32-bit keys per SM per second [31, 32]. For a constant distribution, however, our experiments show an average throughput of only 1.7 billion 32-bit keys per SM per second on an *NVIDIA Titan X (Pascal)*, due to competing updates to only one single shared memory location. This performance drop is shown for the *atomics only* approach in Figure 2, which depicts the memory bandwidth utilisation relative to the peak throughput of 369.17 GB/s (determined using a micro-benchmark for a read-only workload). In contrast, for a uniform distribution over $q$ distinct digit values, with $q \geq 3$, the approach that uses atomics only, sees as much as 3.3 billion updates per SM per second, almost achieving peak memory bandwidth.

In order to avoid such a performance drop for highly skewed distributions, we use the available compute resources for a new approach (*thread reduction & atomics*) that reduces each thread's updates to shared memory. For the simple approach (*atomics only*), the computation for each key is limited to bit-shifting the desired digit to the least-significant digit, masking it, and atomically incrementing the counter for the resulting value in shared memory. Instead, with our improved approach, each thread stores its masked digit values in registers, uses a sorting network to bring them into sorted order and combines the counter updates for subsequent registers sharing the same value into a single *atomicAdd* operation. To limit the complexity of the sorting network, we sort runs of up to nine values at a time using a sorting network that involves 25 comparisons. Once the runs of digit values are in a sorted order, the algorithm iterates over them, combining any sequence of identical digit values into a single *atomicAdd* operation. As shown in Figure 2 (*thread reduction & atomics*), the reduced number of atomic updates now effectively mitigates the performance drop for a very skewed distribution.

Since, the block's histogram needs to be recomputed during the key scattering step, the algorithm stores each block's histogram in device memory to save compute resources later on. This slightly increases the utilised memory bandwidth of this step by a factor of $1 + \frac{r*4}{KPB \times k/8}$, given that the histogram uses counters of four bytes. Assuming a reasonable number of $KPB$, such as 6 912, this adds less than 4% to the data being transferred in the case of 32-bit keys, while saving essential compute resources during the key scattering step.

### 4.4 Key Scattering

For the scattering of a bucket's keys into its $r$ sub-buckets, we use the same subdivision of buckets into key blocks as for the histogram computation. This allows to reuse the histograms that have already been computed and stored in device memory for each block. Each of these histograms indicates the number of keys that are going to be scattered from the key block into each of the sub-buckets. It can therefore be used to determine the size of the chunk of memory within each sub-bucket that needs to be reserved for the block's keys. A chunk of memory for storing $n$ keys within a sub-bucket is reserved by performing a single *atomicAdd* operation that reads the sub-bucket's offset and adds $n$ to it. Adding the number of keys, $n$, to the sub-bucket's offset guarantees that subsequent memory reservations are made beyond this chunk's memory reservation. The original value that has been read before $n$ was added, can therefore be used as the starting offset in memory for the chunk.

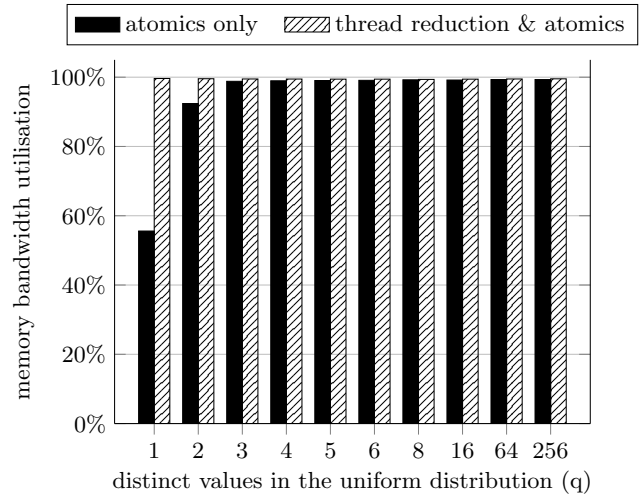

Figure 2: Achieved memory bandwidth utilisation for the histogram computation of a uniform distribution amongst a varying number of values using a non-optimised (atomics only) and an optimised approach (thread reduction & atomics)

Once up to $r$ chunks of memory have been reserved for the block, its keys can be scattered into the reserved memory locations. However, simply scattering the keys to the chunks suffers from irregular memory accesses, as all threads of a thread block write the keys into different chunks residing at distant locations within device memory. To address this issue and coalesce writes to device memory, the keys of each block are first partitioned into the $r$ sub-buckets within shared memory, before writing the whole sub-bucket of a block to the reserved chunk in device memory. Figure 3 illustrates this for a single key block. The top row depicts an excerpt of the device memory holding the input, the middle row represents the local shared memory, and the bottom row shows the device memory for the sub-buckets. The block's keys are read from device memory, partitioned locally into the sub-buckets in shared memory from where the local sub-buckets are finally copied to the chunks that have been reserved within the respective sub-buckets in device memory.

Compared to immediately scattering individual keys to irregular locations in device memory, this considerably improves the memory performance. Yet, depending on the granularity of memory transactions, the choice of $r$ and the number of keys per block, $KPB$, may have considerable implications on the memory efficiency. For memory transac-

422

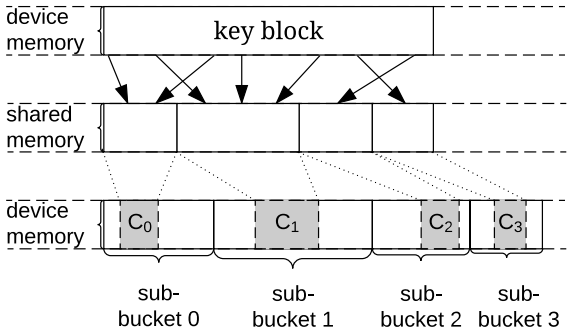

**Figure 3: Using shared memory for write combining**

tions that read or write $T$ bytes at a time, the lower bound of required memory transactions for a block of $k$-bit keys is given by $\lceil \frac{KPB \times k}{T \times 8} \rceil$. That is, for each memory transaction, $T$ bytes are written, with the exception of the last transaction, which possibly only writes the remainder that does not make up $T$ bytes. However, the worst case may require one additional transaction for the remainder of each sub-bucket, totaling $r$ additional memory transactions (neglecting inefficiencies due to misaligned writes). Since the local shared memory is limited to a few tens of kilobytes and has to fit all keys of a block, we are limited to a few thousand keys per block. One possible choice for a key block size would be 32 768 bytes, requiring a minimum of 1 024 transactions for $T = 32$ bytes. Calculating the worst case memory efficiency as the ratio of the lower to the upper bound on the number of memory transactions yields 80% for using eight-bit digits with a radix of 256. Further increasing the digit size to nine, ten, or eleven bits, would further decrease the efficiency to 66.66%, 50%, or 33.33%, respectively. We therefore choose $d = 8$ bits as an optimum trade-off between reducing the number of required sorting passes and the worst case memory efficiency.

The partitioning of a block's keys within shared memory makes use of the shared memory atomics to coordinate writes to the local sub-buckets. Similar to the mechanism being used for reserving chunks within the sub-buckets in device memory, we maintain one write counter in shared memory for each sub-bucket. Prior to writing a key into a local sub-bucket, a thread reads the value from the sub-bucket's write counter and adds the number of keys it intends to write. The original value that is read from the write counter serves as the thread's write offset within the sub-bucket in shared memory. Similar to our histogram approach, this makes extensive use of shared memory atomics. Hence, the key scattering suffers a similar performance drop for skewed distributions as the basic histogram implementation. However, compared to the histogram computation, the key scattering is not limited to just reading the keys from device memory, but also requires writing the keys back, resulting in twice the amount of data being transferred. In order to fully utilise the available memory bandwidth, it is therefore sufficient to achieve only half the processing throughput.

In order to mitigate the performance drop for very skewed distributions, we use an implementation that tries to combine writes of multiple keys to the same local sub-bucket. Instead of writing the keys one by one to the respective local sub-buckets, each thread looks at several keys at a time, writing any sequence of up to three keys sharing the same digit value at once. We refer to this approach as a *look-ahead of two*, since each thread considers the two following keys, in addition to the one it is currently looking at. We chose a look-ahead of two as it provides a reasonable trade-off for maximising the probability of combining writes for the highly skewed distributions, which we are trying to address, without wasting too many compute resources.

In order to avoid the overhead for distributions lacking the skewness to benefit from using a look-ahead due to an insufficiently high probability of finding keys destined for the same sub-bucket, we only consider the look-ahead for highly skewed distributions. Having the block's histogram at hand (from the preceding histogram computation), the algorithm can determine the skewness of the key distribution and only turn to the approach using a *look-ahead* for highly skewed distributions.

### 4.5 Analytical Model

One of the core challenges of the MSD-based hybrid radix sort is that the algorithm may end up with millions and millions of buckets that need to be maintained in memory. This section aims to seize the algorithm's complexity by deducing upper bounds on the maximum number of buckets, blocks, and memory requirements.

The following list presents the most important rules for the sorting algorithm:
(R$_1$) Any bucket of size $n$, with $n \leq \hat{\partial}$, is sorted within on-chip shared memory using a local sort.
(R$_2$) Any bucket of size $n$, with $n > \hat{\partial}$, is partitioned into $r$ sub-buckets using a counting sort.
(R$_3$) Any sequence of sub-buckets is merged as long as the total number of keys falls short of the merge threshold $\underline{\partial}$, with $\underline{\partial} \leq \hat{\partial}$.
(R$_4$) Any bucket of size $n$, with $n > \hat{\partial}$, consists of exactly $\lceil n/KPB \rceil$ blocks and each block holds a sequence of keys from exactly one bucket.

Based on the presented list of rules, the following bounds can be deduced for sorting an input comprised of $n$ keys:
(I$_1$) Following from R$_1$, at any given time, there are at most $\lfloor n/\hat{\partial} \rfloor$ buckets that cannot be sorted with a local sort.
(I$_2$) Following from I$_1$ and R$_2$, at any given time, there are at most a total of $r \times \lfloor n/\hat{\partial} \rfloor$ buckets. This can be deduced, as there are at most $\lfloor n/\hat{\partial} \rfloor$ buckets that are partitioned using a counting sort and each of those buckets is partitioned into at most $r$ sub-buckets.
(I$_3$) Considering R$_3$, the upper bound given by I$_2$ can be refined to $\min(\lfloor 2 \times n/\underline{\partial} \rfloor + \lfloor n/\hat{\partial} \rfloor, r \times \lfloor n/\hat{\partial} \rfloor)$. Following from R$_3$, we conclude that any two subsequent sub-buckets must have at least $\underline{\partial}$ keys, as they would have been merged otherwise. Yet, as we can only merge sub-buckets originating from the same bucket, there may be one sub-bucket per bucket that cannot be merged.
(I$_4$) Following from R$_4$ and I$_1$, at any given time, there are at most $\lfloor n/KPB \rfloor + \lfloor n/\hat{\partial} \rfloor$ blocks. This follows from the fact that there are at most $\lfloor n/KPB \rfloor$ blocks with $KPB$ keys. Adding to that up to one block for the remaining keys of each bucket gives an upper bound on the number of blocks.

Having determined the upper bound on the number of buckets and blocks, the memory requirements can easily be inferred. We are using unsigned integers of four bytes for the counters of the histograms, as well as for the offsets of sub-buckets and key blocks. This can be easily adjusted to support more than $2^{32} - 1$ keys by using a larger data type. For the assignments of thread blocks to key blocks, we are using the following data structure: `{k_offs:uint, k_count:uint, b_id:uint, b_offs:uint}`, holding information on the starting offset of the keys, the number of consecutive keys, the bucket's unique identifier, and its offset. Memory required for these assignments needs to be allocated twice, once to keep track of the assignments of the current pass, and once for the assignments of the subsequent pass. Similarly, we store the following information for the assignment of a bucket whose size falls short of the local sort threshold: `{b_id:uint, b_offs:uint, is_merged:bool}`. In addition to storing one histogram for each bucket exceeding the local sort threshold, we allocate memory for each of its blocks' local histograms. This allows the algorithm to write the local histograms during the histogram computation and reuse the blocks' histograms in the subsequent scattering step.

Apart from the negligible amount of constant memory in the order of a few bytes for the synchronisation between thread blocks, the amount of memory (in bytes) that is required for sorting $n$ keys comprised of $k$ bits is given by:

($M_1$) Input and auxiliary memory: $2 \times n \times k/8$
($M_2$) Bucket histograms: $4 \times r \times \lfloor n/\hat{\partial} \rfloor$
($M_3$) Block histograms: $4 \times r \times (\lfloor n/KPB \rfloor + \lfloor n/\hat{\partial} \rfloor)$
($M_4$) Block assignments: $2 \times 16 \times (\lfloor n/KPB \rfloor + \lfloor n/\hat{\partial} \rfloor)$
($M_5$) Local sort sub-bucket assignments:
  $12 \times \min(\lfloor 2 \times n/\underline{\partial} \rfloor + \lfloor n/\hat{\partial} \rfloor, r \times \lfloor n/\hat{\partial} \rfloor)$

For 32-bit keys, for instance, the total amount of memory required by $M_2$ through $M_5$ is bound by a mere 5% of $M_1$, given a reasonable configuration, such as $KPB = 6\,912$, $\hat{\partial} = 9\,216$, $\underline{\partial} = 3\,000$, and $r = 256$.

## 4.6 Sorting Pairs & Other Data Types

In order to support key-value pairs that are stored in a decomposed layout, the hybrid radix sort is extended to rearrange the values along with the keys they are associated with. Therefore, it is sufficient to adapt the key scattering step and the local sort, which are the only components involved in the permutation of keys. We extend the implementation of the key scattering step to keep track of the memory locations to which the individual keys have been written. Hence, while partitioning a block's keys within shared memory, each thread stores the offsets at which its keys have been placed. Once all keys have been rearranged and the block's local sub-buckets have been copied to device memory, the shared memory can be reused for the values. Each thread reads the values its keys are associated with from device memory and writes them to shared memory according to the offsets that have been stored in the thread's registers during the local partitioning of the keys. Finally, the local sub-buckets holding the values are copied to the respective locations in device memory. The local sort is extended by taking advantage of CUB's *BlockRadixSort* that comes with support for sorting key-value pairs [29]. For key-value pairs that are stored coherently in memory, keys and values need to be decomposed into a key and a value part, recomposing them once the sorting is done. Our experiments have shown that the de- and recomposition can be achieved at peak memory bandwidth, adding only negligible overhead to the sorting procedure.

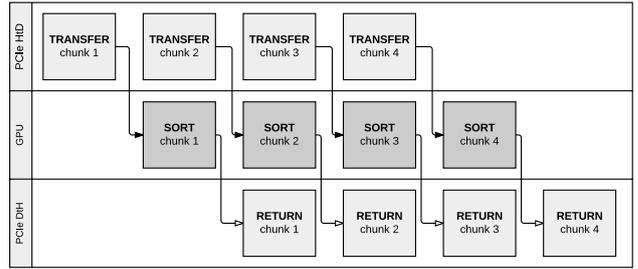

**Figure 4: Pipelined sorting exploiting the available resources to mitigate the data transfer overhead**

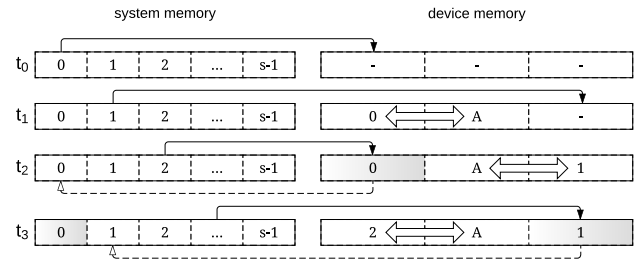

**Figure 5: Efficient device memory utilisation for interleaving sorting with data transfers**

While the presentation of the proposed hybrid radix sort has been limited to sorting unsigned integer keys, it can be easily extended to cover further primitive data types, such as `int`, `float`, and `double`. Support is added by using a bijective mapping from the input's data type to an order-preserving bit-string. This is as simple as flipping the sign-bit for signed integers and a little bit more involved for floats, where all bits have to be flipped if the sign bit was set, and only the sign bit is flipped otherwise [19]. We transform the input during the scattering step of the first counting sort and recover the original representation either during a local sort or the last counting sort pass.

## 5. HETEROGENEOUS SORTING

Having presented an efficient approach for sorting inputs within GPU's device memory, this section builds on that component with a heterogeneous sorting algorithm that addresses inputs that either do not reside on the GPU or simply do not fit into the available device memory. In either case, data has to be transferred over the comparably slow Peripheral Component Interconnect Express (PCIe) bus from the CPU to the GPU and vice versa, adding a considerable amount of overhead to the end-to-end sorting performance. Hence, in addition to the time taken for sorting a given input on the GPU ($T_S$), the time taken for transferring the whole input to the GPU ($T_{HtD}$) as well as the time taken for returning the sorted sequence from the GPU ($T_{DtH}$) have to be considered.

In order to support arbitrarily large inputs and mitigate the overhead that is introduced with the data transfers, we split the input into $s$ chunks and treat them as a set of sub-problems that can be processed concurrently. As illustrated in Figure 4, this allows to overlap the processing stages of multiple sub-problems. For instance, while transferring the data of the third chunk, the GPU can concurrently sort the second chunk and return the sorted run of the first chunk. Since the PCIe bus allows for full-duplex communication, we are able to accelerate data transfers without sacrificing throughput in either direction. With the sorted chunks being returned by the GPU, the CPU is left with the task of merging the $s$ chunks into one final sorted sequence. Denoting the time taken for merging with $T_M$, the end-to-end sorting duration is given by:

$$T_{EtE} = \frac{T_{HtD}}{s} + \max(T_{HtD}, T_S, T_{DtH}) + \frac{T_{DtH}}{s} + T_M$$

Hence, for large enough $s$, the time taken for transferring the input to the GPU, sorting the chunks there and writing the sorted runs back to system memory is now almost down to the time taken for transferring the input over the PCIe bus one single time, or sorting the input on the GPU, whichever takes longer. This carves out a considerable amount of time that the CPU can spend on merging the $s$ chunks. In order to improve the merging performance and avoid being bound by the available memory bandwidth, we use the parallel multiway merge that merges multiple chunks in a single pass from the parallel extension of *stdlibc++*. Moreover, to lower the number of merging passes for larger inputs, we max out the limited device memory with our in-place replacement strategy. That is, rather than allocating memory that can host four chunks: one for sorting, one for the auxiliary memory, one for the chunk being returned from the GPU, and one for copying the next chunk to the GPU, we only require enough memory for three chunks. As depicted in Figure 5 for the first few time-steps, we immediately reuse the memory that is used to hold a sorted chunk by replacing it with the input of the next chunk. At time step $t_2$ in Figure 5, for instance, we return the sorted run for *chunk* 0, while replacing it with the contents of *chunk* 2. This allows supporting larger chunks that may take up almost one third of the available device memory. Assuming a system with sufficient compute power to efficiently merge up to 16 chunks at a time and a GPU with 12 GB of memory, we could sort an input of up to 64 GB using only a single merging pass.

## 6. EXPERIMENTAL EVALUATION

The experiments were conducted on a system running *Ubuntu 16.04* with kernel version 4.4. The system is equipped with 128 GB DRAM (quad-channel, DDR4-2400) and a *Xeon E5-1650 v4* processor with six physical cores, clocked at 3.60 GHz. The source code was compiled with the *O3* flag using release 8.0.44 of the CUDA toolkit. We used driver version 367.48 for an *NVIDIA Titan X (Pascal)* with 12 GB device memory, 3 584 cores, and a base clock of 1 417 MHz. The performance numbers were averaged over 25 runs. We used the CUB header library in version 1.5.1 to compare the presented approach to the state-of-the-art GPU-based radix sort [29]. CUB is developed as an open-source project by NVIDIA Research. The radix sort provided by CUB builds on the approach presented by Merrill et al. [28]. Moreover, we include comparisons to the radix sort implementation of Thrust [20], the merge sort presented by Baxter [4], and the radix sort from Satish et al. [34]. Similarly, we compare the end-to-end sorting performance of our heterogeneous sorting algorithm on the aforementioned system with a six-core CPU to the results that were reported on a stronger system with 32 cores for PARADIS (CPU-based radix sort) [8].

Table 3: Our default configurations

| key/value size | $KPB$ | threads | $KPT$ | $\hat{\partial}$ |
|---|---|---|---|---|
| 32-bit keys | 6 912 | 384 | 18 | 9 216 |
| 64-bit keys | 3 456 | 384 | 9 | 4 224 |
| 32-bit/32-bit pairs | 3 456 | 384 | 18 | 5 760 |
| 64-bit/64-bit pairs | 2 304 | 256 | 9 | 3 840 |

For the counting sort, we used $d = 8$ bits per digit. In order to improve the occupancy, we determined the number of threads as well as the number of keys per thread ($KPT$) based on the amount of shared memory and the number of registers being required by the kernels, which, in turn, depends on the key and value size. Similarly, these factors impose an upper bound on the local sort threshold $\hat{\partial}$, where the kernel's on-chip memory requirements for processing $\hat{\partial}$ elements must not exceed the available resources of a single SM. The values that were determined for these parameters are depicted in Table 3.

Other than comparison-based sorting algorithms, the hybrid radix sort is not prone to the order of the input but rather sensitive to the key distribution. Hence, in order to generate distributions with varying skewness, we implement the benchmark proposed by Thearling et al. [39], which uses the Shannon entropy as a measure of data distribution. Data is generated by repeatedly applying the *bitwise AND* operation to uniform random distributions, which increasingly skews the distribution towards keys with fewer bits set. For 32-bit keys, for instance, an entropy of 32 bits corresponds to a uniform distribution with each single bit of a key having a 50% probability of being set. Repeatedly *AND*ing random keys with such a uniform distribution once, twice, or three times, generates distributions with entropies of 25.96, 17.39, and 10.79 bits, respectively. In order to compare the end-to-end performance to the numbers that have been reported for PARADIS, we also ran experiments with a Zipfian distribution [14, 8].

### 6.1 On-GPU Sorting

We have evaluated the sorting performance for key distributions with varying degrees of skewness, starting from a uniform distribution (32-bit and 64-bit entropy) up to all keys having the same value (zero-bit entropy). Comparing the sorting rates for 32-bit keys (see Figure 6a), the hybrid radix sort shows an improvement of no less than a 1.69-fold speed-up over CUB. Compared to Thrust's radix sort (Thrust), Baxter's merge sort (MGPU), and the radix sort proposed by Satish et al. (Satish et al.), the results show a minimum speed-up of 1.89, 3.96, and 3.66, respectively. Being able to save one sorting pass by finishing early with a local sort, the hybrid radix sort achieves its peak performance for a uniform distribution with more than a two-fold speed-up over CUB, sorting 500 million keys in only 62.6 milliseconds. As shown in Figure 6c, the effect of the local sort becomes even more apparent for 64-bit keys. Sorting a uniformly distributed input of two gigabytes in as little as 66.7 milliseconds, for instance, almost matches the hybrid radix sort's processing duration for 32-bit keys. In contrast, CUB requires roughly twice as many sorting passes for 64-bit keys as for 32-bit keys and therefore sees a 49% performance drop. Starting out with a 3.75-fold speed-up over CUB for

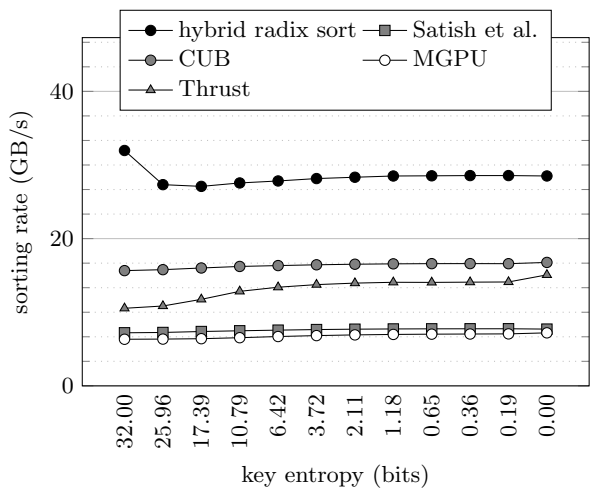 (a) 32-bit keys

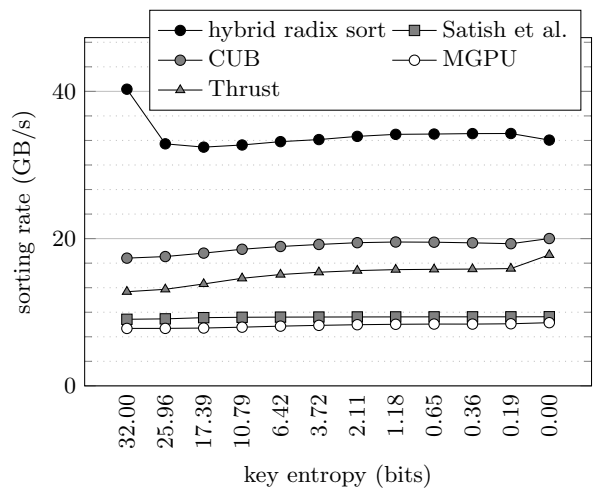 (b) 32-bit keys with 32-bit values

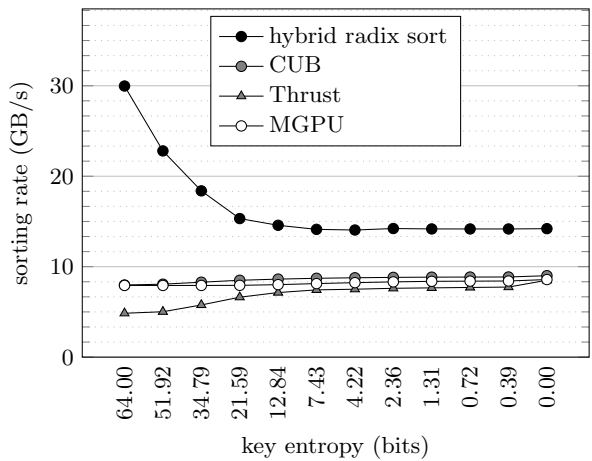 (c) 64-bit keys

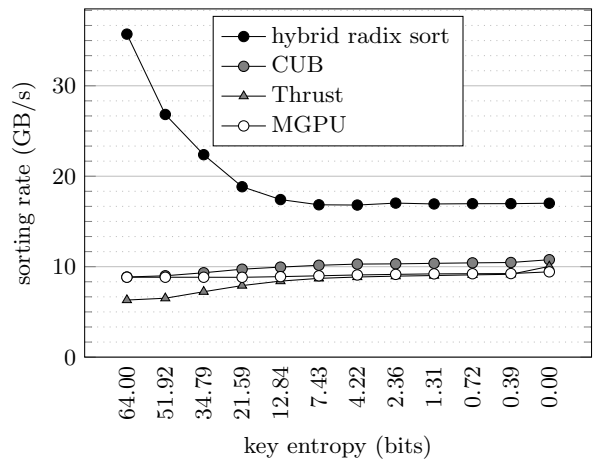 (d) 64-bit keys with 64-bit values

Figure 6: Performance for sorting a 2 GB input with varying data skewness on the GPU

uniformly distributed 64-bit keys, the performance surplus due to the local sort declines for increasingly skewed distributions, flattening out for a distribution with an entropy of zero bits. For such a distribution, all keys have to be run through all counting sort passes. Hence, the performance gain over CUB boils down to the reduced number of counting sort passes and the lower amount of memory transfers. Given keys and key-value pairs that comprise 64-bit keys, an achieved speed-up of the hybrid radix sort with a factor of 1.58 over CUB for such a distribution is in line with the improvements we expect from our 1.625-fold reduction in the amount of memory transfers (13 versus eight sorting passes). Similarly, the 1.7-fold speed-up seen for 32-bit keys closely matches the 1.75-fold improvement over CUB we anticipated as a result of reducing from seven to only four sorting passes. This illustrates that the proposed hybrid radix sort is able to efficiently mitigate the downsides of considering more bits with each sorting pass, achieving more than 97% of the expected theoretical speed-up.

Comparing the hybrid radix sort's performance for sorting key-value pairs to the performance shown for sorting keys only, we see a 20% increase in the amount of data being sorted per second, which matches the reduced amount of memory transfers. Since half the input consists of keys, the hybrid radix sort is reading only half the input during the histogram computation, while still reading and writing the whole input once during the scattering phase. For a total of reading and writing the input only 2.5 times instead of three times, we end up with a 1.2-fold lower amount of memory transfers, which directly translates to a 20% performance increase. This culminates in a sorting rate of up to 40.2 GB/s for 32-bit keys with an associated 32-bit value and up to 35.7 GB/s for 64-bit keys with 64-bit values (see Figure 6b and Figure 6d). Compared to CUB, this corresponds to a 2.32-fold and a four-fold improvement for 32-bit/32-bit key-value pairs and 64-bit/64-bit key-value pairs, respectively.

We also analysed the sorting performance for inputs ranging from 250 000 to 500 million elements with key distributions of varying skewness, i.e., considering an entropy of 64.00, 51.92, 34.79, 21.59, 12.84, 7.43, 4.22, 2.36, 1.31, 0.72,

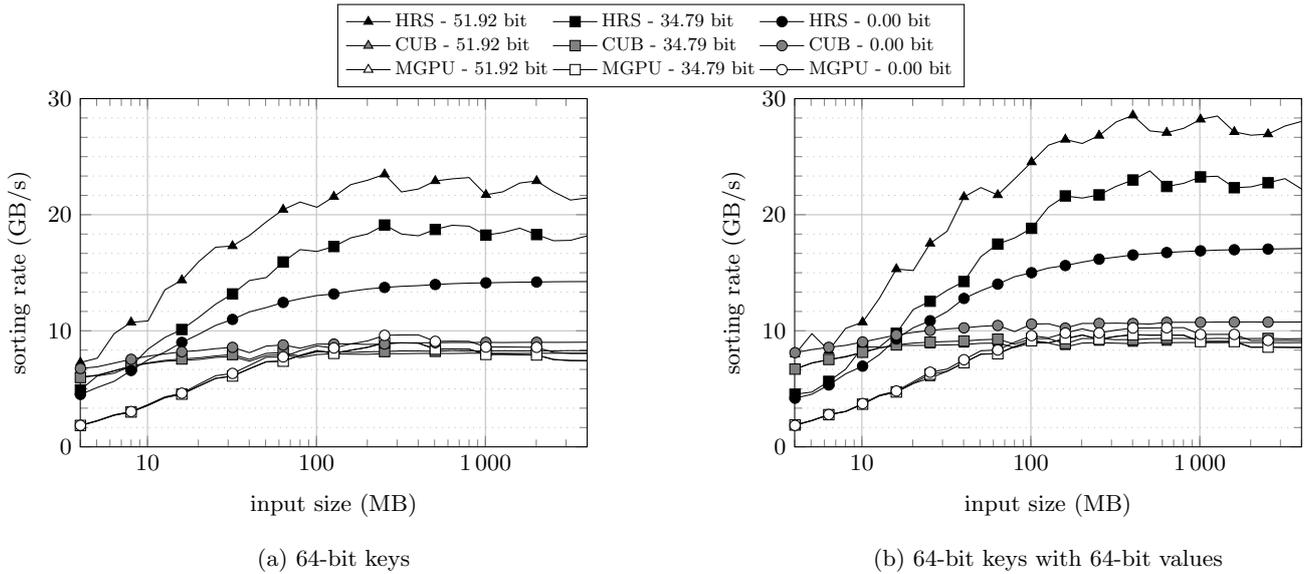

(a) 64-bit keys

(b) 64-bit keys with 64-bit values

Figure 7: Comparison of the hybrid radix sort (HRS), the CUB radix sort (CUB), and merge sort (MGPU) for different distributions with an entropy of 51.92, 34.79, and 0.00 bits

0.39, and 0.00 bits. Being able to save several sorting passes for a uniform key distribution, the hybrid radix sort outperforms CUB for all of the evaluated input sizes. Yet, incurring a slightly lower constant overhead, CUB has an edge for very small and highly skewed inputs that are sorted in the order of hundreds of microseconds (see Figure 7a and Figure 7b). Considering the hybrid radix sort's worst-case key distribution, however, the hybrid radix sort still outperforms CUB for inputs larger than 1.9 million keys and 1.6 million key-value pairs, independently of the key distribution. Given that the input size is a function parameter, we could easily default to CUB's sorting algorithm using a simple case distinction for small inputs that fall short of these thresholds. Compared to Thrust and the GPU-based merge sort (MGPU), our hybrid radix sort is superior for any of the evaluated problem sizes. For reasons of clarity, however, we decided to only present the performance results gathered from the merge sort implementation.

## 6.2 Heterogeneous Sorting

This section analyses the end-to-end sorting performance of the pipelined heterogeneous sorting algorithm and compares it to the numbers reported for the CPU-based radix sort PARADIS [8].

Figure 8 compares the heterogeneous sort to a naïve approach that simply transfers the input to the GPU (*PCIe HtD*), sorts the input there (*on-GPU sorting*), and returns the sorted result over the PCIe bus (*PCIe DtH*). The naïve approach was evaluated for two variants. Firstly, using the state-of-the-art radix sort for the on-GPU sorting (CUB), and secondly, using the hybrid radix sort (HRS). We analysed the performance of the heterogeneous sort for several choices of $s$ (the number of chunks). The figure shows the processing duration of the heterogeneous sort broken down into the *chunked sort* and the *CPU merging*. The *chunked sort* represents the time taken for splitting the input into $s$ chunks, transferring the chunks to the GPU, sorting them on the GPU, and returning the sorted runs over the PCIe bus. The time taken for merging $s$ sorted chunks on a six-core

CPU is depicted by *CPU merging*. Figure 8 shows that, as the number of chunks increases, the time taken by the chunked sort is approaching the time taken for transferring the input one single time over the PCIe bus (cf. Section 5). For $s = 16$ chunks, for instance, the time of the chunked sort is down to 629 milliseconds, which corresponds to a mere 16% more time than it takes to transfer the whole input to the GPU one single time (540 milliseconds). Noticeably, the chunked sort even outperforms the on-GPU sorting time of CUB (636 milliseconds), even though the chunked sort includes the PCIe data transfers to the GPU and back. While we see the performance of the chunked sort improving for a larger number of chunks, our parallel multiway merge lacks the compute power to efficiently merge more than four chunks at a time. For our six-core CPU, we therefore see

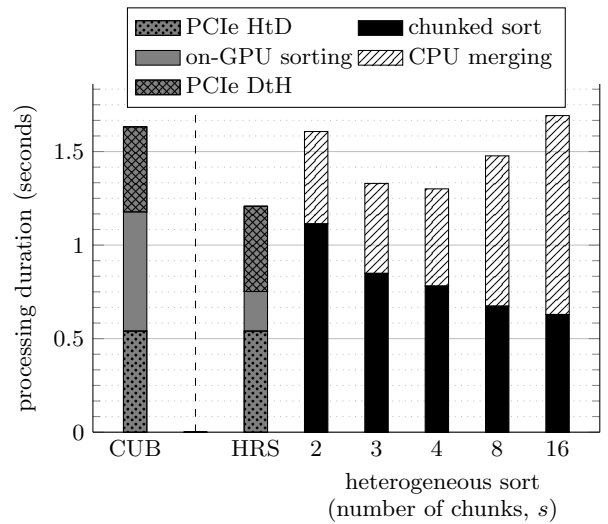

Figure 8: Comparing the end-to-end time for sorting 375 million 64-bit keys with 64-bit values (6 GB)

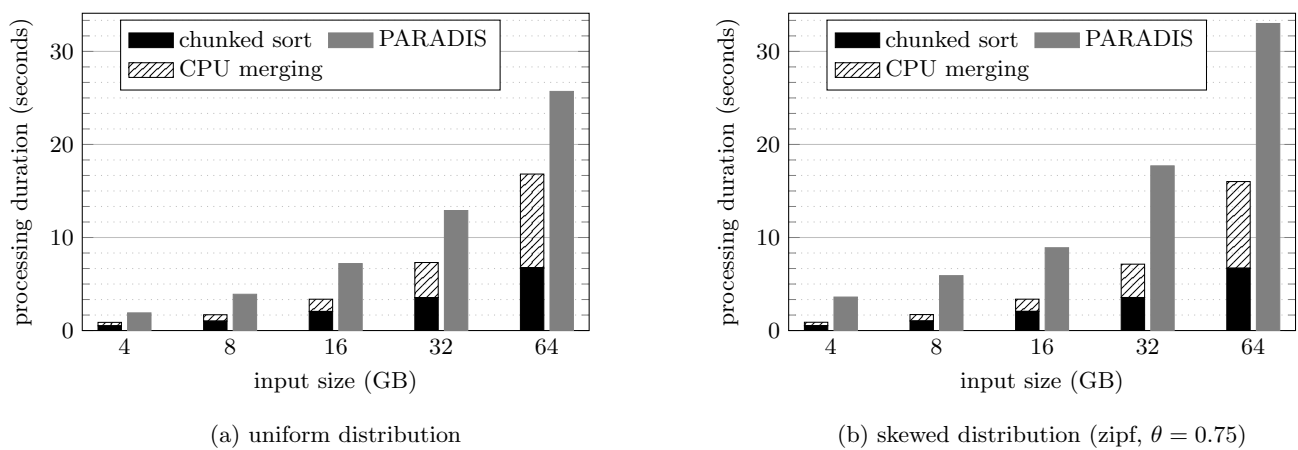

(a) uniform distribution

(b) skewed distribution (zipf, $\theta = 0.75$)

Figure 9: Comparing the end-to-end sorting performance of the heterogeneous sorting algorithm to the state-of-the-art CPU-based radix sort (PARADIS) for inputs comprising 64-bit keys with 64-bit values

a minimum for the overall end-to-end sorting time for four chunks. While these performance numbers are representative for our system, using our merge-based approach, a more powerful host system will see a lower minimum for a higher number of $s$, given that it efficiently merges eight, 16, or even more chunks at a time. Similarly, a more efficient multiway merge implementation or an approach building on partitioning rather than merging may also move the optimum towards a higher number of chunks.

Figure 9a and Figure 9b compare the performance of the heterogeneous sort to the numbers reported for PARADIS running 16 threads on a system with 32 CPU cores [8]. For a skewed distribution, our heterogeneous sorting algorithm achieves a four-fold speed-up, sorting four gigabytes in 895 milliseconds. Even though we see our CPU-based parallel multiway merge slightly degrading the overall performance for larger inputs, the heterogeneous sort still shows more than a two-fold speed-up over PARADIS for an input of 64 GB. While the GPU completes sorting and returning all sorted runs after only 6.7 seconds, it takes the parallel multiway merge on a six-core CPU another 9.3 seconds to merge the sorted runs. Compared to PARADIS, which suffers from skewed distributions, the performance of our approach is almost distribution agnostic, varying by no more than 5% between the uniform and the Zipfian distribution. PARADIS, running 32 threads, takes 19.8 and 25.4 seconds for an input of 64 GB with a uniform and a skewed key distribution, respectively. Even though the heterogeneous sort is only running on a six-core CPU, these results are still up by a factor of 1.18 and 1.59 from the time taken by the heterogeneous sort for a uniform and a skewed distribution, respectively.

## 7. CONCLUSIONS

This work presented a novel approach to radix sorting on GPUs. Instead of building on the common LSD radix sort approach for GPUs that relies on stable sorting passes, we took a different route with our efficient implementation of an MSD radix sort. Proceeding from the most-significant to the least-significant digit allows our algorithm to drop the requirement of stable sorting passes. By lifting this constraint, we were able to substantially reduce the number of required sorting passes and the amount of memory transfers. For the memory bandwidth-bound radix sort, we achieve a baseline of a 1.6-fold reduction in the amount of memory transfers, which directly translates to an achieved minimum speed-up of a factor of 1.58. This shows that our approach is successfully addressing the challenges arising from implementing an MSD radix sort on GPUs, such as load balancing issues for skewed distributions and performance degradation due to bucket handling, while still being able to max out the high memory bandwidth of GPUs. Moreover, sorting small buckets in on-chip memory rather than running them through subsequent partitioning passes enables additional performance improvements, culminating in a four-fold speed-up over the state-of-the-art approach.

In addition, we presented a heterogeneous sorting algorithm that uses the CPU on powerful host systems to mitigate the overhead introduced with PCIe data transfers and sort arbitrarily large inputs. Using pipelining, we were able to exploit the full-duplex communication of the PCIe bus, while interleaving the process of sorting and data transfers. Transferring an input to the GPU, sorting it into runs of up to four gigabytes each, and returning the sorted runs is now almost as fast (i.e., 9.55 GB/s) as transferring the input in one direction, one single time over the PCIe bus (i.e., 12 GB/s). Comparing the end-to-end sorting performance of our heterogeneous sort (including the time taken for merging the runs on a six core CPU) to the numbers reported for PARADIS using 16 threads on a 32 core system, we see a 2.2-fold and a four-fold speed-up for an input of four gigabytes with a uniform and a Zipfian distribution, respectively. Even though being bound by the merging performance of the CPU for larger inputs, like 64 GB, we still see an improvement of a factor of 1.52 and 2.07 for a uniform and a Zipfian distribution, respectively.

## 8. ACKNOWLEDGMENTS

This research has been supported by the Alexander von Humboldt Foundation. We would also like to thank Saman Ashkiani and the other authors of GPU Multisplit for sharing their implementation with us.

# APPENDIX
## A. ADDENDUM ON THE LATEST WORK

As a fundamental operation that finds its application in many fields, GPU-based sorting algorithms receive a lot of attention. Given the strong interest in efficient sorting algorithms, available implementations are continuously improved and new approaches are regularly published. With this addendum we aim to meet the rapid advancements that are made in this field, covering up to date work, which followed our initial submission and the completion of the peer review process, with preliminary and non-exhaustive results that we were able to obtain just in time with the authors' support. In particular, that is the work of Ashkiani et al., who present an improved version of their multisplit primitive (GPU Multisplit) that can be used for the partitioning passes of a radix sort as well as an update of the CUB library, which in version *1.6.4* enables specific GPU architectures to support up to seven bits per sorting pass [2, 29]. While CUB is maxing out shared memory at the cost of lower occupancy, GPU Multisplit makes use of the warp-synchronous execution and warp-wide intrinsics for the efficient data exchange between threads of the same warp to mitigate excessive on-chip memory requirements.

Figure 10 shows a performance comparison of the hybrid radix sort and the two latest approaches, putting their advancements into context by adding the prior state-of-the-art baseline (CUB, version 1.5.1) to the evaluation. For sorting 32-bit keys, the hybrid radix sort still achieves as much as a 56% improvement over CUB's latest version. For any non-constant distribution, it retains a minimum improvement of 32% over CUB (version 1.6.4), with an edge of 21% for a constant distribution (0 bits entropy). For 32-bit keys, GPU Multisplit is superior to CUB (version 1.5.1), yet, inferior to CUB (version 1.6.4). The hybrid radix sort outperforms GPU Multisplit by no less than a factor of 1.53 for 32-bit keys (see Figure 10a). As shown in Figure 10b, GPU Multisplit and CUB in its latest version are roughly on a par for sorting key-value pairs (32-bit keys with 32-bit values). While GPU Multisplit has an edge over CUB of up to 12%

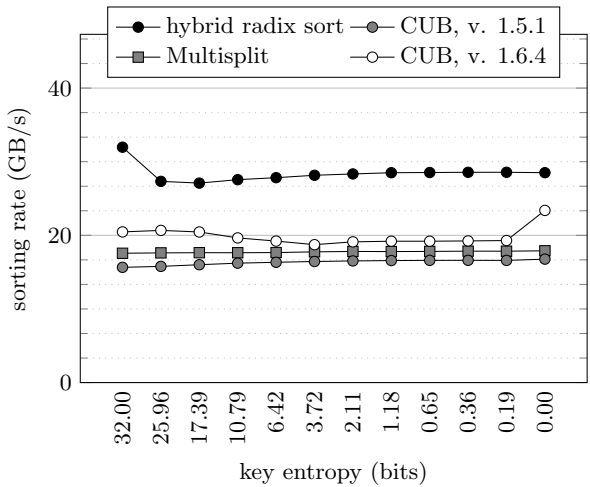
(a) 32-bit keys

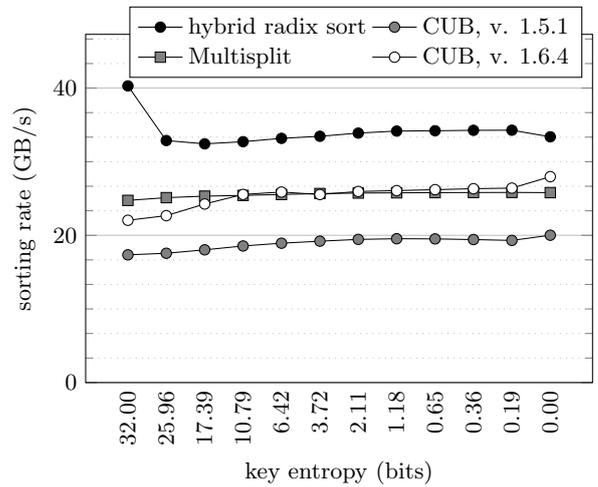
(b) 32-bit keys with 32-bit values

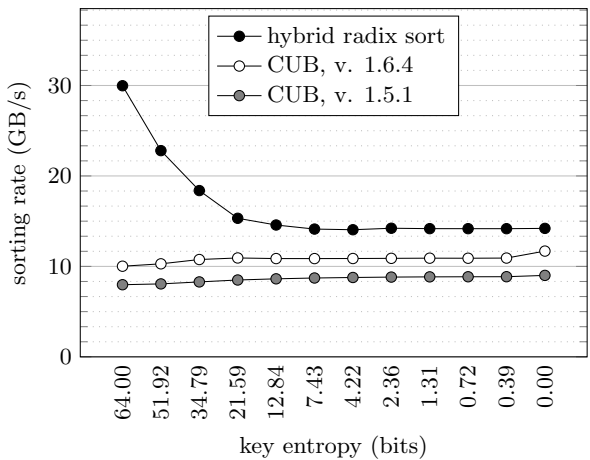
(c) 64-bit keys

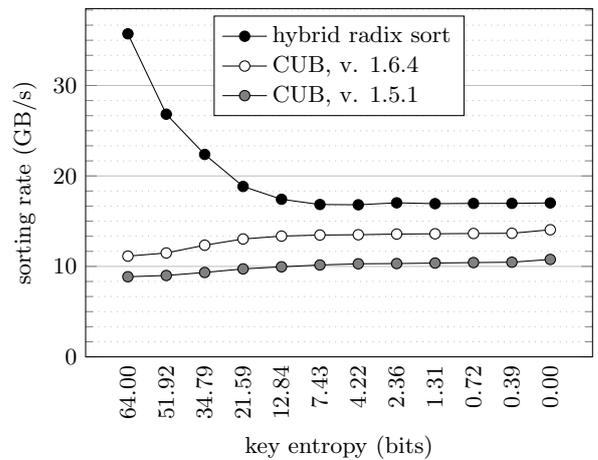
(d) 64-bit keys with 64-bit values

Figure 10: Performance for sorting a 2 GB input with varying data skewness on the GPU

for more uniform distributions, CUB (version 1.6.4) has an edge of up to 8% for highly skewed distributions. Compared to GPU Multisplit, the hybrid radix sort achieves as much as a 1.62-fold improvement, with a minimum speed-up of 1.29. Compared to CUB's latest version, the hybrid radix sort shows an improvement of up to 82% and no less than 28% for any non-constant distribution. Similarly, the hybrid radix sort provides a minimum speed-up over CUB (version 1.6.4) of 1.29 for 64-bit keys over any non-constant distribution, showing as much as a 2.99-fold improvement for a uniform distribution (see Figure 10c). For key-value pairs (64-bit keys with 64-bit values), the hybrid radix sort outperforms CUB (version 1.6.4) by a factor of 3.21 for a uniform distribution, while still showing no less than a 21% improvement for any of the remaining distributions (see Figure 10d).

## B. IMPACT OF OPTIMIZATIONS

While building on an MSD-based hybrid radix sort enables the performance improvements with considerable speed-ups in the first place, it also makes the algorithm highly sensitive to the input distribution. To ensure that the algorithm provides relatively constant performance results, even for challenging input distributions that are highly skewed or that would otherwise require handling millions and millions of buckets, this work has developed several optimisations. In order to show the impact of individual optimisations, we rerun our experiments with single optimisations being switched off. For our evaluation, we distinguish between independent optimisations that are analysed by disabling them individually and a group of synergistic optimisations. The performance impact of disabling a combination of independent optimisations can easily be approximated by multiplying the relative performance impact of the individual optimisations. Disabling a combination of optimisations within the group of synergistic optimisations (i.e., *single local sort config* and *no bucket merging*), in contrast, may have a more drastic effect than their multiplicative performance impact, since the lack of one optimisation may boost the impact of the absence of the other optimisation. Therefore, in addition to switching off individual optimisations within the group, we also evaluated the performance impact for disabling the combination of synergistic optimisations.

For the group of synergistic optimisations, our analysis

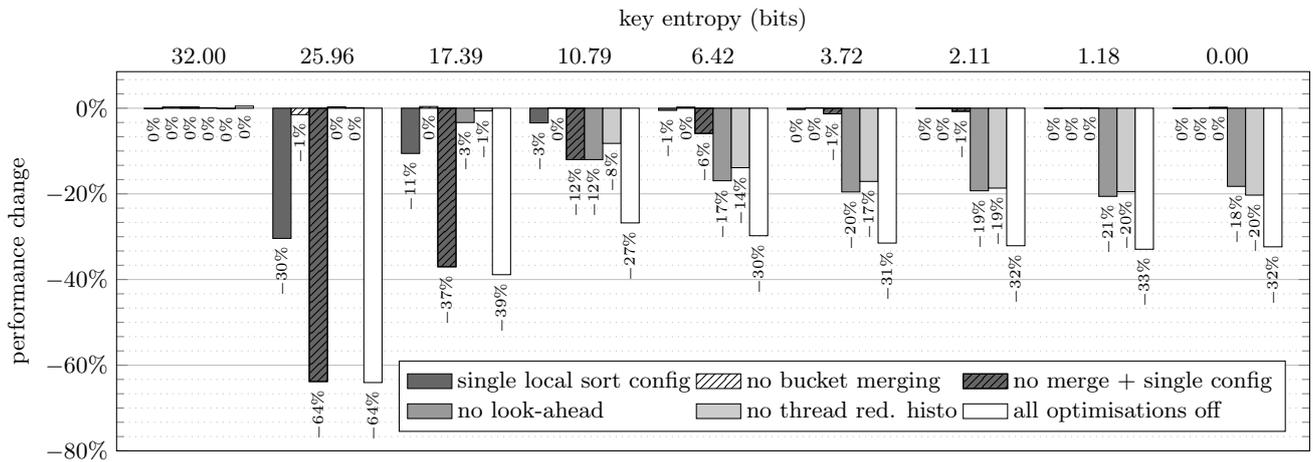

Figure 11: Performance impact on the sorting rate of 32-bit keys, when switching off individual optimisations

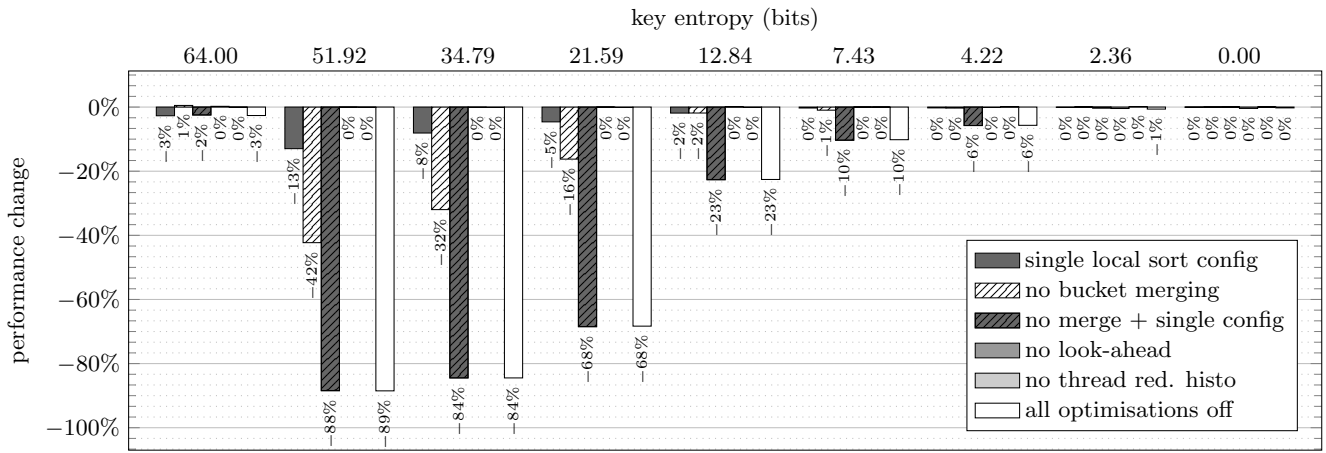

Figure 12: Performance impact on the sorting rate of 64-bit keys, when switching off individual optimisations

431

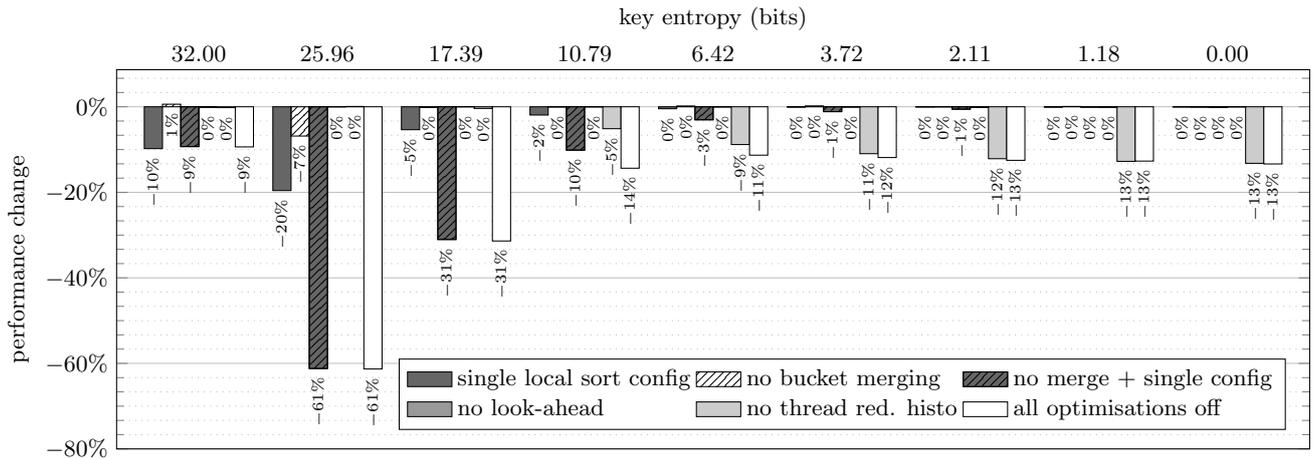

Figure 13: Performance impact on the sorting rate of 32-bit keys with 32-bit values, when switching off individual optimisations

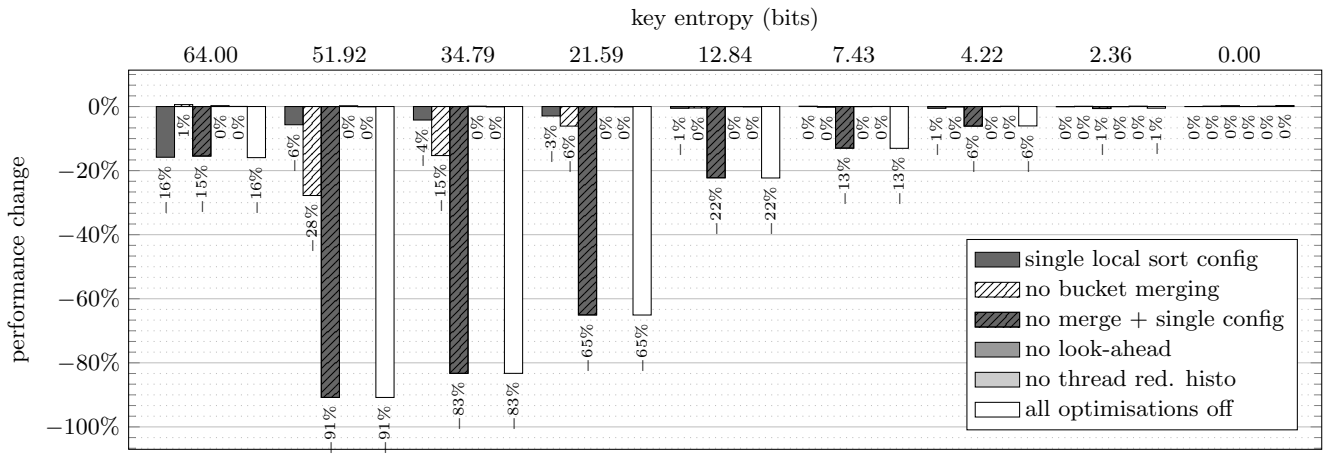

Figure 14: Performance impact on the sorting rate of 64-bit keys with 64-bit values, when switching off individual optimisations

considers using only a single local sort configuration (*single local sort config*) that sorts any bucket of up to $\hat{\partial}$ keys, not merging tiny buckets (*no bucket merging*), as well as the combination of both (*no merge + single config*). Amongst the independent optimisations, we considered not using the look-ahead during the scattering step (*no look-ahead*) and not using the thread reductions during the histogram computation (*no thread red. histo*).

Figure 11 shows the performance impact of switching off individual optimisations when sorting 32-bit keys. The performance impact is depicted as a performance delta, with the percentage denoting the performance increase or drop, after switching off an optimisation, compared to the performance achieved with all optimisations in place. Similarly, Figure 12, Figure 13, and Figure 14, depict the same information for sorting 64-bit keys, 32-bit keys with 32-bit values, and 64-bit keys with 64-bit values, respectively, showing the performance impact of individual optimisations.